\documentstyle[12pt,aaspp]{article}
\newcommand{\beq}{\begin{equation}}
\newcommand{\eeq}{\end{equation}}
\newcommand{\ombcc}{1-\beta_{\parallel}\cos\theta}
\newcommand{\dc}{\Delta\!\chi}
\newcommand{\NarrowMargins}{
  \setlength{\oddsidemargin}{+0.3in}
  \setlength{\evensidemargin}{-0.0in}
  \setlength{\textwidth}{6.2in}
  \setlength{\topmargin}{-0.75in}
  \setlength{\textheight}{9.25in}   }
\catcode`\@=11 
\def\lsim{\mathrel{\mathpalette\@versim<}}
\def\gsim{\mathrel{\mathpalette\@versim>}}
\def\@versim#1#2{\vcenter{\offinterlineskip
        \ialign{$\m@th#1\hfil##\hfil$\crcr#2\crcr\sim\crcr } }}
\catcode`\@=12 
\NarrowMargins
\begin{document}
\title{Harmony in Electrons:  Cyclotron and Synchrotron Emission \\
by Thermal Electrons in a Magnetic Field.} 
\author{Rohan Mahadevan, Ramesh Narayan, and Insu Yi
\footnote{Institute for Advanced Study, Princeton, NJ 08540.}}
\affil{Harvard-Smithsonian Center for Astrophysics,
60 Garden St., Cambridge, MA 02138.}

\begin{abstract}

We present a complete solution to the cyclotron-synchrotron radiation
due to an isotropic distribution of electrons moving in a magnetic
field.  We make no approximations in the calculations other than
artificially broadening the harmonics by a small amount in order to
facilitate the numerics.  In contrast to previous calculations, we sum
over all relevant harmonics and integrate over all particle and
observer angles relative to the magnetic field.  We present emission
spectra for electron temperatures $T=5\times10^8$ K, $10^9$ K,
$2\times10^9$ K to $3.2\times10^{10}$ K, and provide simple fitting
formulae which give a fairly accurate representation of the detailed
results.  For $T\geq3.2\times10^{10}$ K, the spectrum is represented
well by the asymptotic synchrotron formula, which is obtained by
assuming that the radiating electrons have Lorentz factors large
compared to unity.  We give an improved fitting formula also for this asymptotic
case.

\end{abstract}

\section{Introduction}

Radiation by quasi-relativistic and fully relativistic electrons in
magnetic fields is very common in astrophysics.  Examples of sources
include supernova remnants (e.g. Anderson, Keohane \& Rudnick 1995),
jets and lobes in radio galaxies (e.g. Carilli {\it et al.} 1991), hot
accretion flows onto neutron stars and black holes (e.g. Narayan, Yi
\& Mahadevan 1995), and relativistic fireballs in gamma-ray bursts
(e.g. Paczy\'{n}ski \& Rhoads 1993, Meszaros, Laguna \& Rees 1993).
In all these examples we have thermal or non-thermal electrons
radiating cyclo-synchrotron radiation in a magnetic field which is
probably of near-equipartition strength.  It is clearly important to
be able to calculate the radiative luminosities and spectra of these
systems.

Surprisingly, a full solution to this problem does not appear to have
been published so far, especially for thermal sources.  The majority
of previous work (e.g. Schott 1912, Schwinger 1949, Oster 1960, 1961,
Pacholczyk 1970) is limited either to the cyclotron or the synchrotron
limit where analytical formulae may be used.  These calculations are
not relevant in the transition zone between cyclotron and synchrotron
radiation when the velocities of the electrons are quasirelativistic.
Petrosian (1981) has obtained some approximate formulae in this
intermediate regime, but a complete analysis of the emission from
thermal sources with temperatures in the range $10^9-10^{10}$ K can
only be done numerically.  Takahara \& Tsuruta (1982) and Melia (1994)
have reported numerical calculations in the mildly relativistic
regime, but both calculations involve some approximations and do not
agree with each other.  Therefore, there is need for a more accurate
treatment to resolve this difference.  In
particular, we find significant differences between the results quoted
by Melia (1994) and the exact calculations presented here.

We present in this paper the complete solution to the
cyclo-synchrotron problem for a thermal plasma.  In \S \ref{theory} we
give a qualitative as well as quantitative description of the problem,
and briefly discuss previous work in this subject.  We also present an
outline of how we solve the problem exactly.  Then in \S \ref{results}
we present the results.  We calculate the emission due to a thermal
distribution of electrons for a range of temperatures, and provide
analytic fitting functions for the spectra.

\section{Theory}
\label{theory}

\subsection{Qualitative Features}

A non-relativistic particle moving in a magnetic field radiates with a
simple dipole pattern (e.g. Rybicki \& Lightman 1979) so that an
observer sees a sinusoidal electric field.  The observed spectrum
consists of a single delta function at the orbital frequency,
$\omega_o$, of the electron.  This is the limit of extreme cyclotron
radiation.  If the particle is speeded up, the dipole radiation
pattern gets distorted and the observed electric field is no longer
purely sinusoidal.  The spectrum then picks up additional Fourier
components, or harmonics, which are integral multiples of the
fundamental frequency.  The situation is illustrated in Fig. 1a.  At
the same time, $\omega_o$ decreases with increasing velocity $v$,
according to
\beq
\omega_o = {\omega_b \over \gamma}, \mbox{\hspace{2cm}} 
\omega_b = {e B \over m_e c}, 
\eeq
where $\omega_b$ is the cyclotron frequency of the particle, and
$\gamma = 1/\sqrt{1-v^2/c^2} = 1/\sqrt{1-\beta^2}$ is the Lorentz
factor.  

With increasing $\gamma$, the emission is beamed more and more towards
the direction of motion of the particle.  For large $\gamma$, the
electric field appears as a series of delta function-like pulses, each
of width $\sim 1/\gamma^2 \omega_b$, with succeeding pulses separated
by a time $2\pi \gamma /\omega_b$.  The Fourier transform of these
delta functions gives closely spaced delta functions in frequency
which approximate a nearly continuous spectrum extending up to $\omega
\sim \gamma^2\omega_b$, and cutting off exponentially at higher
$\omega$.  This corresponds to the synchrotron limit, where the
spectrum has the universal shape shown in Fig. 1b.

The cyclotron and synchrotron limits have very different spectra, as
Fig. 1 shows.  How does the spectrum evolve from one to the other as
the particle velocity is increased?  Answering this question requires
detailed calculations and is the topic of this paper.

\subsection{Previous Work} 

We begin by reviewing a number of well-known results.  The power
(energy/time/steradian/frequency) emitted by an electron moving with a
velocity parameter between $\vec{\beta}$ and $\vec{\beta} +
d\vec{\beta}$, in a frequency range $d\omega$, and at an observer
angle $\theta$ with respect to the magnetic field, is given by
(e.g. Schott 1912, Rosner 1958, Bekefi 1966)
\beq
\eta_{\omega}(\vec{\beta},\theta) \, d\omega= {e^2 \omega^2 \over 2 \pi c}
	\left[ \sum\limits_{m=1}^{\infty} \left( {\cos\theta - 
\beta_{\parallel} \over \sin\theta} \right)^2 \, J^2_m(x) + \beta^2_{\perp}
J^{\prime 2}_m(x) \right] \delta(y) \, d\omega. \label{eta}
\eeq
Here 
\beq
x = {\omega \over \omega_o} \beta_{\perp} \sin \theta, \label{x} 
\eeq
\beq
y = m\omega_o - \omega(1-\beta_{\parallel}\cos\theta), \label{y}
\eeq
$\delta(y)$ is the Dirac delta function, $J_m(x)$ is the Bessel
function of order $m$, $J^{\prime}_m(x)$ is its derivative,
$\beta_{\parallel} = \beta \cos\theta_p$ is the velocity parameter
parallel to the magnetic field $B$, where the particle moves at an
angle $\theta_p$ to the local field, and $\beta_{\perp} = \beta
\sin\theta_p$ is the velocity parameter perpendicular to the magnetic
field.

Each integer $m$ in the summation in equation (\ref{eta}) corresponds to a
harmonic, and the presence of the $\delta$-function implies that the
emission occurs at discrete frequencies.  To calculate the power in
successive harmonics at low $\beta$ (the cyclotron limit), we expand
the Bessel functions to lowest order in $x$.  After integrating over
all observer angles, the emission in each harmonic is given by (Schwinger 1949, Rosner
1956, Bekefi 1966)
\beq
\eta_m^T =  {2e^2 \omega^2_b \over c} {(m+1)(m^{2m+1}) \over
                                (2m+1)!} \beta^{2m},
\eeq
This result, which is valid so long as $m\beta \ll 1$, shows that the
spectrum decreases rapidly with increasing $m$.  Fig. 1a shows an
example with $\beta = 0.1$, which corresponds to this limit.

In the opposite limit of $\gamma\gg1$, which is the synchrotron limit,
we are interested in large $m$'s, and the Bessel functions can be
approximated by modified Bessel functions (Bekefi 1966). We then have
the familiar result (e.g. Schwinger 1949, Pacholczyk 1970, Rybicki
\& Lightman 1979)
\beq
{dE \over d\omega}  =  {\sqrt{3} e^3 B \sin\theta_p \over 2\pi m_e c^2} F(X),
\label{syncformula}
\eeq
where
\beq
X = {\omega \over \omega_c}, \mbox{\hspace{1cm}} \omega_c = { 3\over 2} \, 
	\gamma^2 \omega_b \sin\theta_p,
\eeq
$F(X)$ is given by
\beq
F(X) \equiv X \int\limits_X^{\infty} K_{{5\over 3}}(\xi)\, d\xi, 
\eeq
and $ K_{{5\over 3}}(\xi)$ is the modified Bessel function.  By taking
appropriate limits of $F(X)$, we obtain 
\begin{eqnarray}
F(X) &\rightarrow&  { 4\pi \over \sqrt{3} \, \Gamma\left( {1 \over 3}\right)}
	\, \left( {X \over 2} \right)^{1/3}, \mbox{\hspace{1cm} $X \ll 1$}, 
\label{foxlx} \\
F(X) &\rightarrow& \left( {\pi \over 2} \right)^{1/2} e^{-X} X^{1/2}, 
\label{foxhx}
\mbox{\hspace{1cm} $X \gg 1$.}
\end{eqnarray}
Fig. 1b shows the synchrotron spectrum for a particle with $\gamma =
10$.

The above results are for electrons of a given velocity.  For a
thermal distribution we need to integrate the emission over a
Maxwellian.  This integral can be done in the synchrotron limit to
give (Pacholczyk 1970),
\beq
\varepsilon_{\omega} d\omega= C \, {\chi \over K_2(1/\theta_e)} \,  
I\left( {x_M \over \sin\theta_p} \right) \, d\omega
	\mbox{\hspace{.3cm}ergs s$^{-1}$ Hz$^{-1}$}, \label{pasyn}
\eeq
where
\beq
C \equiv {e^2 \, n_e \, \omega_b \over \sqrt{3} \, \pi c}, 
\mbox{\hspace{1cm}} \chi \equiv {\omega \over \omega_b}, \mbox{\hspace{1cm}} 
x_M \equiv { 2 \, \chi \over 3 \, \theta_e^2}, \mbox{\hspace{1cm}} \theta_e 
\equiv {kT \over m_e c^2},
\eeq 
$K_2(x)$ is the modified Bessel function, and $I(x_M)$ is defined by
\beq
I(x_M) \equiv {1 \over x_M} \int\limits_0^{\infty} z^2 \exp(-z) \, F(x_M/z^2) dz.
\label{ioxeq}
\eeq
In writing equation (\ref{pasyn}), we have included the proper
normalization for a relativistic Maxwellian distribution; in the limit
of high temperatures, this normalization is equivalent to that used by
Pacholczyk (1970). The limiting behavior of $I(x_M)$ for small and
large $x_M$ are straightforward.  The former has been worked out by
Pacholczyk (see Eq. (\ref{a1})), and the latter by Petrosian (1981) (see
Eq. (\ref{eqna4}). Equation (\ref{pasyn}) corresponds to particles at a
fixed angle $\theta_p$ to the magnetic field.  For an isotropic
particle distribution, we need to perform another integral over
$d(\cos\theta_p)$.  The asymptotic dependencies of the result for small
and large $x_M$ are again easy to calculate and are discussed in
Appendix A.

For the mildly relativistic case, Petrosian (1981) has obtained the
following analytic approximations for the emission spectrum from a
thermal distribution of electrons as a function of the observer angle
$\theta$:
\beq
\varepsilon_{\omega}(\theta) \, d\omega \rightarrow
{2 \pi \over 6^{1/2}} \, C \, {\chi \over K_2(1/\theta_e)}  
\, \exp \left( - 6.75^{1/3} 
\, x_M^{1/3} \right) \, d\omega ,  
\mbox{\hspace{.3cm} $\chi \, \theta_e \gg 1$}, \label{pet1}
\eeq 
and 
\begin{eqnarray}
\varepsilon_{\omega}(\theta) \, d\omega &\rightarrow&
3^{1/2} \, 2^{3/2} \pi^2  \, C \, {1 \over \theta_e \, K_2(1/\theta_e)} 
\left( {\chi \, \theta_e} \right)^{3/2} \nonumber \\ 
&\times& \left(
{ 1 + \cos^2\theta \over \sin^2\theta } \right) \, \exp \left[
- \chi \ln\left( {2\, \over e \, \chi \theta_e \sin^2\theta } \right) \right]
\, d\omega, \mbox{\hspace{.3cm} $\theta_e \ll 1$}. \label{pet2}
\end{eqnarray}
These results are valid only with the additional condition
$\omega/\omega_b \gg 1$, the regime Petrosian (1981) was interested
in.  For $x_M \gg 1$, equation (\ref{pet1}) is exactly the same as
equation (\ref{pasyn}) (cf. Eq. \ref{eqna4}), and we see that the
approximations made by Petrosian (1981) in this regime are the same as
the extreme synchrotron approximations.

The extreme synchrotron results described above provide quite an
accurate representation of the exact results for electron temperatures
$\gsim 3\times 10^{10}$K, while the series result given in equation
(5) is very good at temperatures $\lsim 10^8$K.  For $10^8$K $\lsim T
\lsim 3 \times 10^{10}$K, we cannot use either of these limiting
results but must solve the complete cyclo-synchrotron problem.

\subsection{The Complete Solution}

We start by writing the integrals that need to be carried out.  To
determine the luminosity, $L_{\omega} \equiv dE/d\omega$, due to
a particle moving with velocity parameter $\beta$, we must integrate
$\eta_{\omega}(\vec{\beta}, \theta)$ (Eq. (\ref{eta})) over observer
angles, and a given particle distribution:
\beq
L_{\omega} \equiv { dE \over d\omega}  = {2 \over 4 \pi}
	\int\limits_0^1 d\beta \, n(\beta) 
	\int\limits_0^{2\pi} d\phi_p \int\limits_0^1 d(\cos\theta_p) 
	\int\limits_0^{2\pi} d\phi   \int\limits_{-1}^1 d(\cos\theta)
	\, \eta_{\omega}(\vec{\beta}, \theta), \label{fulleq}
\eeq
where $\eta_{\omega}(\vec{\beta}, \theta)$ is given by
Eq. (\ref{eta}).  The subscript $p$ refers to the particle, and
$n(\beta)$ is the velocity distribution of the particles, which is
taken to be isotropic.  The factor of $2$ in front is because we
integrate over only half the range of $\cos\theta_p$, and the factor
of $1/4\pi$ comes from the angular normalization of the (isotropic)
particle distribution function.  The integrals over $\phi_p$ and
$\phi$ are trivial, giving
\beq
L_{\omega} \equiv { dE \over d\omega}  = 2\pi \int\limits_0^1
	 d\beta \, n(\beta)
        \int\limits_0^1 d(\cos\theta_p)
        \int\limits_{-1}^1 d(\cos\theta)
        \, \eta_{\omega}(\vec{\beta}, \theta). \label{L_w}
\eeq
For a fixed velocity parameter $\beta$ and frequency $\omega$, we
numerically evaluate the two innermost integrals in Eq. (\ref{L_w}),
as well as the sum over harmonics in the expression (2) for
$\eta_{\omega}(\vec{\beta}, \theta)$.  We repeat this calculation for
various values of $\beta$ and $\omega$ and tabulate the results.  The
results can then be convolved with any isotropic velocity distribution
to obtain the spectrum $L_{\omega}$.  In this paper, we restrict
ourselves to a relativistic Maxwellian $n(\beta)$ and present
detailed results for this particular case.

The evaluation of Eq. (\ref{L_w}) involves a delta function
(Eq. (\ref{eta})), which determines the precise frequencies at which
radiation is observed.  For a given harmonic $m$, the delta function
implies that there is emission only at
\beq
  \omega = {m \omega_o \over (1 - \beta_{\parallel}\cos\theta)}. \label{fmold} 
\eeq 
Since $m$ takes on only integer values, this means that for a given
$\vec{\beta}$ and $\cos\theta$, emission is observed only at a
discrete set of $\omega$.  Unfortunately, the discrete nature of the
emission poses a serious numerical difficulty since it requires
infinite resolution in frequency.  In order to make the numerics
tractable, we replace the delta function with a smooth broadening
function, $f(\omega)$, which is nonzero over a finite frequency range
$\omega_c \pm \Delta \omega$.  Here $\omega_c$ is the central
frequency where we wish to evaluate the emission, and $\Delta \omega$
is a broadening width, which we adjust.  With the smoothing function,
the delta function in equation (2) becomes
\beq
\delta(y) =  \delta \left[ m \omega_o - \omega(1 - \beta_{\parallel}\cos\theta) \right]
= {1 \over \omega_b} \, {1 \over 1 - \beta_{\parallel}\cos\theta} \,
	f(\chi), \label{deltf}
\eeq
where 
\beq
\chi = {\omega \over \omega_b} = {\omega \over \gamma \omega_o }. \nonumber
\eeq
For $f(\chi)$, we choose the functional form
\beq
f(\chi) =  {15 \over 16 \Delta\!\chi} \left[ 1 - \left( 
{2 \over \Delta\!\chi^2} \right) (\chi-\chi_c)^2 + \left( {1 \over \Delta\!\chi^4}
\right) (\chi-\chi_c)^4 \right], \label{fochieq}
\eeq
which has the property that $f(\omega_c \pm \Delta \omega) =
f^{\prime}(\omega_c \pm \Delta \omega) = 0$.

In our calculations we set $\dc = \alpha \chi$, and we have found that
a choice $\alpha = 0.05$ allows the delta function to be broadened
sufficiently to stabilize the numerics without losing too many details
of the harmonic structure of the emission.  For each value of $\chi_c$
of interest, we calculate the power emitted into a width $2\dc$
centered at $\chi_c$, counting only those harmonics which fall within
this width.  With Eqs. (\ref{eta}), and (\ref{deltf}),
Eq. (\ref{L_w}) thus becomes
\begin{eqnarray}
&L_{\omega}& = 2\pi \left( {e^2 \omega_b \over 2 \pi c} \right)
 \int\limits_0^1 d\beta \, n(\beta)
        \int\limits_0^1 d(\cos\theta_p)
        \int\limits_{-1}^1 d(\cos\theta)
        \, {f(\chi^{\prime}) \over 1 - \beta_{\parallel}\cos\theta} 
\nonumber  \\
	&\times& \, \chi^{\prime 2} \left[ \sum\limits_{m=1}^
		{\infty} \left( {\cos\theta -
\beta_{\parallel} \over \sin\theta} \right)^2 \, 
J^2_m\left( \gamma \chi^{\prime} {\beta_{\perp} \sin\theta \over \ombcc}
\right) + \beta^2_{\perp} J^{\prime 2}_m
\left( \gamma \chi^{\prime} {\beta_{\perp} \sin\theta \over \ombcc}\right) \right].  \label{maineq}
\end{eqnarray}
For a Maxwellian velocity distribution with temperature $T$ , we have
\beq
        n(\gamma) \, d\gamma = {m_e c^2 \over k T K_2(1/\theta_e)}
                                \  \beta \gamma^2 \exp(-\gamma/\theta_e)
                                d\gamma , 
\eeq
or
\beq
        n(\beta) \, d\beta = {m_e c^2 \over k T K_2(1/\theta_e)}
                                \ \beta^2 \gamma^5 \exp(-\gamma/\theta_e)
                                d\beta ,
\eeq
with
\beq
        \int_1^{\infty} n(\gamma) \, d\gamma = \int_0^1 n(\beta) \, d\beta
                = 1 .
\eeq
Here $k$ is Boltzmann's constant, $\theta_e = kT/m_ec^2$, and $K_2(x)
$ is the modified Bessel function.

The procedure to calculate the full cyclo-synchrotron emission
spectrum is as follows.  Using Eq. (\ref{maineq}), we fix the values
of $\beta $, $\cos\theta_p$ , $\cos\theta$, and sum over the harmonics
responsible for emission over the frequency range $\chi-\dc$ to
$\chi+\dc$.  This gives the total emission into the given observer
angle $\theta$ from particles at angle $\theta_p$.  We then integrate
over all observer ($\cos\theta$) directions, to get the emission into
all of observer space, and integrate over all particle
($\cos\theta_p$) directions to obtain the total emission per particle
at the given value of $\chi$ due to an isotropic distribution of
particles moving with the given $\beta$ or Lorentz $\gamma$.  We
repeat the calculation for various values of $\gamma$ and $\chi$ to
obtain a table of values.
 
The present calculations differ from previous work by Tsuruta
\& Takahara (1982) and Melia (1994) in that we
integrate over all observer angles whereas those authors restricted
themselves to a particular observer angle.

\section{Results}
\label{results}

\subsection{Transition from Cyclotron to Synchrotron.} 
\label{re_de}

Figure 2 shows plots of the emission as a function of scaled frequency
$\chi$ for various particle velocities, after the angular integrations
in Eq. (\ref{maineq}) have been performed.  For comparison, the
dashed lines show the synchrotron formula as given in
Eq. (\ref{syncformula}).  We see that the exact numerical results
deviate considerably from the limiting synchrotron formula in panels
(a)-(c), where the harmonic cyclotron-like emission is quite evident.
The harmonics are broadened, partly because of our broadening function
$f(\chi)$ (Eq. (\ref{fochieq}), and partly due to Doppler broadening
induced by $\beta_{\parallel}$ (see Eq. (\ref{fmold})).  The former
effect dominates in panel (a), but the latter is more important in
panel (c).

By panel (d), which corresponds to $\beta=0.9, \ \gamma = 2.3$, the
harmonics have merged completely at higher values of $\chi$ and the
synchrotron approximation is becoming quite good.  Yet higher values
of $\gamma$ (panels (e)-(f)) make this trend more apparent.  However,
at low values of $\chi$ the harmonics continue to be present and there
are deviations from the synchrotron formula.

Figs. 3a-d show polar plots of the emission as a function of the
observer angle $\theta$.  In these figures, the particle has $\gamma =
10.3$ and is moving in a helix at an angle $\theta_p$ to the magnetic
field with $\cos\theta_p = 0.3$. The field points along the positive
y-axis.  Fig. 3d shows the emission pattern that is observed at a
large frequency $\chi = 100$.  According to Fig. 2f, which corresponds
to this value of $\gamma$, the emission at this $\chi$ is practically
equal to the synchrotron formula.  The reason is clear from Fig. 3d.
We see that the emission is tightly beamed along the particle
direction, and the beam has a half angle $\sim 1/\gamma$, exactly as
assumed when deriving the synchrotron formula.  Varying $\cos\theta_p$
does not change this behavior and, therefore, on integrating over
$\cos\theta_p$ we find a total emission which agrees very well with
the synchrotron formula.

In strong contrast is the case shown in Fig. 3a, which corresponds to
the same values of $\gamma = 10.3$ and $\cos\theta_p= 0.3$, but has a
much lower value of $\chi=0.8$.  The emission pattern now is not
strongly beamed about the particle's velocity.  Instead we can clearly
see a set of harmonics, each with a width determined by $f(\chi)$.  As
$\cos\theta_p$ varies, the individual beams get longer or shorter, but
the harmonic character remains.  Therefore, the integrated emission
over all $\cos\theta_p$ deviates significantly from the synchrotron
approximation (as Fig. 2f shows at $\log(\chi) \simeq -0.1$).
Figs. 3b, c, show how the transition in the beaming occurs as $\chi$
increases.

\subsection{Emission by a Thermal Distribution of Electrons}

The primary output from our calculations is a table of $\eta_{\chi}$
values as a function of electron velocity $\beta$ (or equivalently
$\gamma$) and dimensionless frequency $\chi$.  This table can be
convolved with any isotropic velocity distribution to calculate the
corresponding spectrum.  In the following we describe the results for
a thermal Maxwellian distribution.

\subsubsection{The Ultra-relativistic Regime}
\label{ultraregime}

We first consider the ultra-relativistic regime, where $kT \gg
m_ec^2$.  Pacholczyk (1970) derived the formulas given by
Eqs. (\ref{pasyn}) and (\ref{fulleq}), for the emission from a
relativistic Maxwellian distribution of electrons with a fixed
particle angle $\theta_p$.  For an isotropic distribution of
particles, we define a new function $I^{\prime}(x_M)$
\beq
I^{\prime}(x_M) = { 1 \over 4\pi} \int I\left( {x_M \over
        \sin\theta_p} \right) \, d\Omega_p.
\eeq
and substitute $I^{\prime}(x_M)$ for $I(x_M)$ in Eq. (\ref{pasyn}).

Pacholczyk (1970) has numerically calculated $I(x_M)$ over a wide
range of $x_M$ and tabulated the values.  In addition he has shown
that in the limit of small $x_M$
\beq
I(x_M) \rightarrow {1 \over x_M} \, { 16 \pi \over 9 \sqrt{3}} \left(
	{x_M \over 2} \right)^{1/3} \simeq 2.5593 x_M^{-2/3}.
\eeq
In the opposite limit of $x_M \gg 1$, Petrosian (1981) finds
\beq
I(x_M) \rightarrow 2.5651 \exp(-1.8899 \, x_M^{1/3}).
\eeq
We can combine these two limiting extremes into the following simple
fitting function,
\beq 
I(x_M) = 2.5651 \left( 1 + {1.92 \over x_M^{1/3}} + {0.9977 \over
 x_M^{2/3}} \right) \, \exp(-1.8899 \, x_M^{1/3}), \label{eqnIxM}
\eeq
where we have optimized the coefficient $1.92$ in the middle term so
as to minimize the error.  This fitting function has a maximum error
of $0.39\%$, which occurs when $x_M \approx 63$.  Figures 4a, b,
compare the fitting function with the exact numerical values of
$I(x_M)$ and show the residuals.

In the case of $I^{\prime}(x_M)$, we show in Appendix
\ref{ioxappendix} that as $x_M \rightarrow 0$
\beq
I^{\prime}(x_M) \rightarrow 2.1532 \ x_M^{-2/3},
\eeq
and as $x_M \rightarrow \infty$
\beq
I^{\prime}(x_M) \rightarrow 4.0505 \  x_M^{-1/6} \exp(-1.8899 \,x_M^{1/3}).
\eeq
Once again we combine these two limits to obtain a fitting function:
\beq 
I^{\prime}(x_M) = {4.0505 \over x_M^{1/6}} \left( 1 + { 0.40 \over
        x_M^{1/4}} + {0.5316 \over x_M^{1/2}} \right) \exp \left(
        -1.8899 \, x_M^{1/3} \right).
\label{ip1}
\eeq
This function has a maximum error of $2.7\%$ at $x_M \approx 160$.
Figs. 4c, d, compare the fitting function with the numerical values
and show the residuals.

\subsubsection{The Mildly Relativistic Regime}

Using our tabulated values of $\eta_{\chi}$ as a function of $\beta$
and $\chi$, we have computed emission spectra for isotropic thermal
distributions of electrons with temperatures in the range, $5\times
10^8$K $< T < 3.2 \times 10^{10}$K.  
We see that at the highest temperatures the emission is very
similar to $I^{\prime}(x_M)$, equation (\ref{ip1}), 
but there are significant deviations at
lower temperatures, especially at small $\chi$.  We have obtained a
set of fitting functions corresponding to each of the temperatures for
which we have calculated the spectrum.  Each function is of the form
\beq
M(x_M)  = {4.0505 \alpha \over x_M^{1/6}} 
	\left( 1 +  {0.40 \beta \over x_M^{1/4}} +
       {0.5316 \gamma \over x_M^{1/2}} \right) 
	\exp \left( -1.8896 \, x_M^{1/3} \right), \label{meq}
\eeq
where $\alpha, \, \beta, \, \gamma$, are all adjustable parameters
which we have optimized so as to minimize the square of the deviation
of $M(x_M)$ from the numerically calculated results.  Table
\ref{taberrwor} shows the optimized parameters we obtained at the various 
temperatures.  We expect that as $T\rightarrow \infty$, $\alpha$,
$\beta$, and $\gamma$ should all~$\rightarrow 1$, since $M(x_M)$ must
approach $I^{\prime}(x_M)$.  Indeed we see that this is the case in
Table \ref{taberrwor}.  At lower values of $T$, however, the
parameters are very different from 1.  This is to a large extent
because the fitting function is attempting to fit the harmonic
``bumps'' in the spectrum

In terms of the fitting function $M(x_M)$, the optically thin
cyclo-synchrotron emission from a thermal plasma at temperature $T$
is given by (cf. equation (\ref{pasyn})),
\beq
\varepsilon_{\omega} d\omega= C \, R(\chi, T) \, d\omega
        \mbox{\hspace{.3cm} ergs s$^{-1}$ Hz$^{-1}$}, \label{mny2}
\eeq
with 
\beq
R(\chi,T) = {\chi \, M(x_M) \over K_2(1/\theta_e)},
\eeq
and the power is given by
\beq
\omega \, \varepsilon_{\omega} = \chi \, L_{\chi} 
= \omega_b \, C \, \chi R(\chi, T)
 \mbox{\hspace{.3cm} ergs s$^{-1}$}. \label{mny3}
\eeq
For temperatures $ T>3\times10^{10}$ K, $M(x_M)$ should be replaced by
$I'(x_M)$ given in equation (\ref{ip1}), which is equivalent to setting the 
fitting constants $\alpha, \ \beta, \ \gamma$ equal to unity.

Figs. 5, 6, show the fits at various temperatures along with the
residuals, and Table \ref{table2wor} shows where the maximum errors
occur, both over the range $\log(\chi) <1$ and for $\log(\chi) >1$.
We see that the errors are particularly severe for $\log(\chi) < 1$
because of the harmonic oscillations which are impossible to fit in
detail with a simple function such as (\ref{meq}).  However, in many
applications, the synchrotron emission will be self-absorbed, and one
would be interested primarily in $\log(\chi) >1$.  We see that the
errors here are much less severe.

We have compared our detailed numerical results with those of
Petrosian (1981), Takahara \& Tsuruta (1982) and Melia (1994).  We
find good agreement with the former two papers.  The small differences
in our results can be explained by the fact that they only considered
a single observer direction whereas we have averaged over all
directions.  We do, however, find a serious discrepancy with Melia's
results.  For instance, our calculation shows that for $T = 10^{10}$K,
the spectrum peaks at $\log(\chi) \simeq 2$, whereas the calculation
by Melia indicates that the peak is at $\log(\chi)\simeq 1$ and that
the emission is insignificant at $\log(\chi)\simeq 2$.  In fact,
Melia's results seem to indicate that the cutoff frequency for the
synchrotron emission is essentially independent of the temperature,
whereas it is clear from basic principles that the cutoff must
increase rapidly with increasing temperature. 

\section{Conclusion}
\label{conclusion}
The detailed calculations described in this paper bridge the gap
between the limits of nonrelativistic cyclotron emission and
ultrarelativistic synchrotron emission.  We have calculated the
emission due to an isotropic distribution of charged particles moving
in a magnetic field, and have shown how the spectrum changes as a
function of the particle Lorentz factor $\gamma$ and the dimensionless
frequency $\chi$ (defined in Eqs. 20 and 1).  Figure 2 shows some
results for selected cases.  Included in our calculations are all the
details of the harmonic emission.  This is important at low values of
$\gamma$, and for low frequencies even at high values of $\gamma$.
Also, we isotropically average over observer directions relative to
the magnetic field.

Having calculated emission spectra for an array of values of $\gamma$,
we have integrated the spectra over an isotropic relativistic
Maxwellian distribution of particle velocities to calculate the
spectrum of cyclo-synchrotron emission due to a thermal plasma in a
magnetic field.  The results are shown in Figs. 5 and 6 for
temperatures ranging from $5\times10^8$ K to $3.2\times10^{10}$ K.  We
have obtained fitting functions $M(x_M)$ with three fitting constants,
$\alpha$, $\beta$, $\gamma$ (see Eqs. 33--35 and Table 1), which
provide a fairly accurate representation of the numerical results.
These fitting functions allow the spectrum to be calculated with
reasonable accuracy for any temperature $>5\times10^8$ K.  The errors
decrease with increasing temperature in the manner indicated in Table
2.

The thermal cyclo-synchrotron spectra presented in this paper agree
with most previous results, except that our calculations are more
complete since the spectra have been integrated over all particle and
observer angles.  At highly relativistic temperatures, our results
agree with those given by Pacholczyk (1970), while at mildly
relativistic temperatures and for frequencies $\omega \gg \omega_b$,
our results agree with those of Petrosian (1981) and Takahara \&
Tsuruta (1982).  There are minor deviations in the results which can
be traced to the fact that the previous authors specified a fixed
direction of the observer relative to the field rather than averaging
over all directions.  Our results do, however, differ significantly
from the calculations presented by Melia (1994) and we have been
unable to understand the reason for the discrepancy.

Finally, we note that the basic output of our calculations is a table
of emission spectra for isotropic particles of fixed velocity $\beta$
or Lorentz factor $\gamma$.  In this paper we have concentrate on one
application of this table, namely the calculation of thermal
cyclo-synchrotron spectra from thermal plasmas with Maxwellian
velocity distributions.  The tabulated results could be convolved with
any other isotropic electron distribution function, e.g. a power law
distribution, to calculate the corresponding spectrum.  Our work thus
provides a ``ready to use'' table for determining the
cyclo-synchrotron emission from any astrophysical source with an
isotropic particle and magnetic field distribution.

\noindent
Acknowledgements: 
This work was supported in part by NSF grants AST 9423209 (to the
Center for Astrophysics) and PHY 9407194 (to ITP, University of
California, Santa Barbara).  RN thanks the ITP for hospitality.

\pagebreak
\begin{appendix}
\section{Asymptotic  Formulae for $I(x)$ and $I^{\prime}(x)$}
\label{ioxappendix}

\subsection{$I(x)$}

The definition of $I(x)$ is given in Eq. (\ref{ioxeq}):
\beq
I(x) \equiv {1 \over x} \int\limits_0^{\infty} z^2 \exp(-z) \,
F(x/z^2) dz.
\eeq
By using the approximation of $F(x)$ given in Eq. (\ref{foxlx}),
Pacholczyk (1970) showed that,
\beq
I(x) \rightarrow {1 \over x} \, { 16 \pi \over 9 \sqrt{3}} \left( {x
        \over 2} \right)^{1/3} \simeq 2.5593 x^{-2/3},
        \mbox{\hspace{1cm} $x \ll 1$.}\label{a1}
\eeq
In the opposite limit of $x\gg1$, we use the approximation of $F(x)$
given in Eq. (\ref{foxhx}) to write
\beq
I(x) \rightarrow \sqrt{{\pi \over 2}} \, {1 \over x} \, x^{1/2}
	\int\limits_0^{\infty} z^2 {1 \over z} \exp[-(z + x/z^2)] \,
	dz.
\eeq
Employing the method of steepest descent, Petrosian (1981) showed that this 
integral can be evaluated to give 
\begin{eqnarray}
I(x) &\rightarrow& {2 \pi \over 6 ^{1/2}} \exp[-(2^{1/3} + 2^{-2/3})\,
        x^{1/3}], \nonumber \\  &\rightarrow& 2.5651 \exp(-1.8899 \,
        x^{1/3}), \mbox{\hspace{1cm} $x \gg 1$.} \label{eqna4}
\end{eqnarray}

\subsection{$I^{\prime}(x)$}

The function $I^{\prime}(x)$ is defined by
\beq 
I^{\prime}(x) = { 1 \over 4\pi} \int I\left( {x \over
        \sin\theta_p} \right) \, d\Omega_p.
\eeq
We first consider the case for $x \ll 1$, where we have $I(x)
\rightarrow 2.56x^{-2/3}$.  Setting $x \rightarrow x/\sin\theta$,
we obtain
\beq
I^{\prime}(x) \rightarrow 2 \cdot {2\pi \over 4\pi} 
\int\limits_0^1 {2.56 \over x^{2/3}} 
\, \sin^{2/3}\theta \, d(\cos\theta), 
\eeq
which can be evaluated to give
\begin{eqnarray}
I^{\prime}(x) &\rightarrow& {2.56 \over x^{2/3}} \, {\Gamma(1/2)
		\Gamma(1/3) \over 2 \, \Gamma(11/6)}, \\ \nonumber
		&\simeq& 2.153 x^{-2/3}, \mbox{\hspace{1cm} $x \ll
		1$.}
\end{eqnarray}

For $x \gg 1$, we have $I(x) \rightarrow 2.5651\exp(-1.8899x^{1/3})$,
and setting $x \rightarrow x/\sin\theta$, we obtain
\beq
I^{\prime}(x) = 2.5651 \int\limits_0^1
\exp(-1.8899x^{1/3}/\sin^{1/3}\theta) d(\cos\theta).
\eeq
We now use the fact that most of the emission comes from $\theta \simeq
\pi/2$.  Setting $\theta = \phi + \pi/2$ we obtain
\beq
I^{\prime}(x) \rightarrow 2.5651\int\limits_0^1 \exp(-1.8899x^{1/3}/
\cos^{1/3}\phi) \, d(\sin\phi).
\eeq
Setting $\cos\phi = (1-y^2)^{1/2}$, and Taylor-expanding up to $y^2$,
we finally obtain
\begin{eqnarray}
I^{\prime}(x) &\rightarrow& 2.5651 \int\limits_0^1
\exp(-1.8899x^{1/3}) \,
\exp(-1.8899x^{1/3} \, y^2/6) \, dy, \\ \nonumber
&=& {2.5651 \exp(-1.8899x^{1/3}) \over 2} \sqrt{{\pi \over 1.8899}} \, 
\sqrt{{6 \over x^{1/3}}}, \\ \nonumber
&\simeq& 4.05047 { 1\over x^{1/6}} \exp(-1.8899x^{1/3}),
\mbox{\hspace{1cm} $x \gg 1$.}
\end{eqnarray}

\end{appendix}
\clearpage
\pagestyle{empty}
\pagebreak
\begin{table}[h] 
\caption[taberrwor]{Optimal values of the 
parameters for different temperatures.}
\begin{center} 
\begin{tabular}{cccc}\hline \label{taberrwor}
$T \ (K)$ & 
	\multicolumn{3}{c}{$I^{\prime}(x)$} \\ \hline
& $\alpha$ & $\beta$ & $\gamma$ \\ \hline
$ 5 \times 10^8$ & 0.0431 & 10.44 & 16.61 \\
$ 1 \times 10^9$ & 1.121 & -10.65 & 9.169 \\
$ 2 \times 10^9$ & 1.180 & -4.008 & 1.559 \\
$4 \times 10^9 $ & 1.045 & -0.1897 & 0.0595 \\
$ 8 \times 10^9$ & 0.9774 & 1.160 & 0.2641 \\
$ 1.6 \times 10^{10}$ & 0.9768 & 1.095 & 0.8332 \\
$ 3.2 \times 10^{10}$ & 0.9788 & 1.021 & 1.031 \\ \hline
\end{tabular}
\end{center}
\end{table}
\clearpage
\begin{table}[h]
\caption[table2wor]{List of errors and 
where they occur for $\log(\chi) <1$ and for
$\log(\chi) > 1$, for different temperatures.}
\begin{center} 
\begin{tabular}{ccccc} \hline \label{table2wor}
$T \ (K)$ & $\log(\chi) < 1$ & \% Error & $\log(\chi) > 1 $& \% Error \\
&  $\log(\chi_{max. error})$ & & $\log(\chi_{max. error})$ & \\ \hline
$ 5 \times 10^8$ & 0.1 & $ 440 $ & 2.3 & 85.1 \\
$ 1 \times 10^9$ & 0.05 & 33.6 & 3.3 & 16.6 \\
$ 2 \times 10^9$ & 0.0 & 5.9 & 3.0 & 5.3 \\
$4 \times 10^9 $ & 0.33 & 6.7 & 2.8 & 2.4 \\
$ 8 \times 10^9$ & 0.33 & 5.3 & 1.9 & 0.9\\
$ 1.6 \times 10^{10}$ & 0.0 & 7.0 & 1.4 & 0.56 \\
$ 3.2 \times 10^{10}$ & 0.0 & 7.4 & 1.6 & 0.55 \\ \hline
\end{tabular}
\end{center}
\end{table}

\clearpage
\noindent {\large \bf References} \\

 \noindent Anderson, M. C., Keohane, J. W., Rudnick, L., 1995, ApJ, 441, 300\\
 Bekefi, G., 1966, Radiation Processes in Plasmas (New York: John Wiley
\& Sons, Inc.)\\
 Carilli, C. L., Perley, R. A., Dreher, J. W., Leahy, J. P., 1991, ApJ, 383, 
554\\
 Melia, F., 1994, ApJ, 426, 577\\
 M\'{e}sz\'{a}ros, P., Laguna, P., Rees, M. J.,  1993, ApJ, 415, 181 \\
 Narayan, R., Yi, I., Mahadevan, R., 1995, Nature, 374,623 \\ 
 Oster, L., Phys. Rev., 1960, 119, 1444\\
 Oster, L., Phys. Rev., 1961, 121, 961\\
 Pacholczyk, A. G., 1970, Radio Astrophysics (San Francisco: Freeman)\\
 Paczy\'{n}ski, B., Rhoads, J. E., 1993, ApJ, 418, L5 \\
 Petrosian, V., 1981, ApJ, 251, 727\\
 Rosner, H., 1958, Republic Aviation Corporation, Missile 
Systems Division, Technical Report No. 206-950-3, ASTIA AD 208 852 \\
 Rybicki, G., Lightman, A., 1979, Radiative Process in 
Astrophysics (New York: John Wiley \& Sons, Inc.)\\
 Schott, G. A., 1912, Electromagnetic Radiation (London: Cambridge University Press)\\
 Schwinger, J., 1949, Phys. Rev., 75, 1912\\
 Takahara, F., Tsuruta, S., 1982, Progress of Theoretical Physics, 
Vol. 67, No.2, 485 \\

\clearpage
\newpage
\noindent{\bf Figure Captions} \\
\noindent Figure 1:  (a) Cyclotron emission  for a particle with $\beta = 0.1$.
The emission is normalized to the total emission in all harmonics.
(b) Synchrotron emission for a particle with $\gamma = 10$ from
standard synchrotron theory. \\

\noindent Figure 2:  Plots of scaled emission $\log(l_{\chi})$ against 
$\log(\chi)$ for (a)
$\beta = 0.2$, (b) $\beta= 0.3$, (c) $\beta= 0.6$, (d)$\beta=0.9$,
(e)
$\gamma = 5.3$, (f) $\gamma= 10.3$, (g) $\gamma= 40.3$, (h)$\gamma=100.3$. 
The vertical axis is the scaled emission, $l_{\chi}$ 
after the angular integrations
in Eq. (\ref{maineq}) have been performed.  
$l_{\chi} \equiv L_{\chi} \, (c/e^2 \, \omega_b^2) = 
L_{\omega} \, (c/e^2 \, \omega_b)$ (cf. Eq.(\ref{maineq})). The
dashed lines  show the synchrotron limit given by Eqn. (\ref{syncformula}). \\

\noindent Figure 3:  Polar plot of emission from a particle moving with a
Lorentz factor $\gamma = 10$ at an angle of $\cos\theta_p = 0.3$.
The magnetic field is oriented along the positive y-axis and $\cos\theta = 0$
 corresponds to the x-axis. The values of $\chi$ are  (a) $\chi = 0.8$, (b)
 $\chi = 1.4$, (c) $\chi = 5.0$, (d) $\chi = 100.0$. \\

\noindent Figure 4:  (a) Plot of the analytic and numerical values for $I(x)$ and
(b) the corresponding residuals.
(c) Plot of the analytic and numerical values for
$I^{\prime}(x)$ and (d) the
corresponding residuals. \\

\noindent Figure 5:  (a) Plots of Eqn. (\ref{mny2}) (dashed lines)  and the numerical
calculation (solid lines) for
(a) $T = 3.2\times10^{10}$, (b) $T = 1.6\times10^{10}$, (e)
$T = 8\times10^{9}$, (f) $T = 4\times10^{10}$, and their corresponding
percentage errors (c), (d), (g), and (h). \\

\noindent Figure 6:  (a) Plots of Eqn. (\ref{mny2}) (dotted lines)  
and the numerical
calculation (solid lines) for
$T = 2\times10^9$, (b) $T = 1\times10^{9}$, (e) $T = 5\times10^{8}$, and their
corresponding percentage errors (c), (d), and (f).

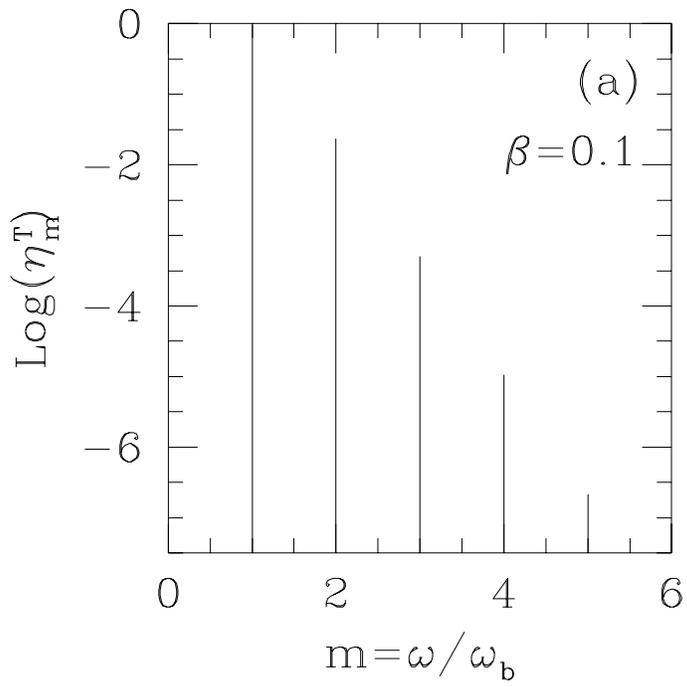
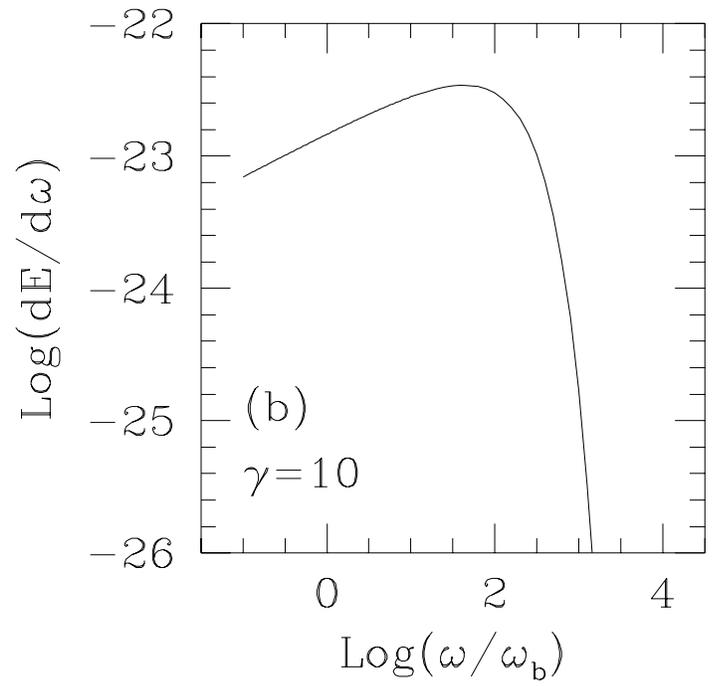
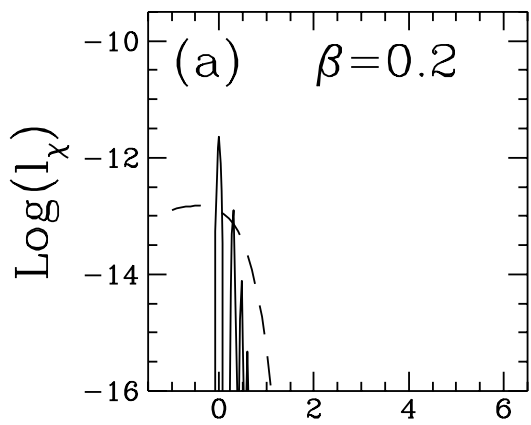
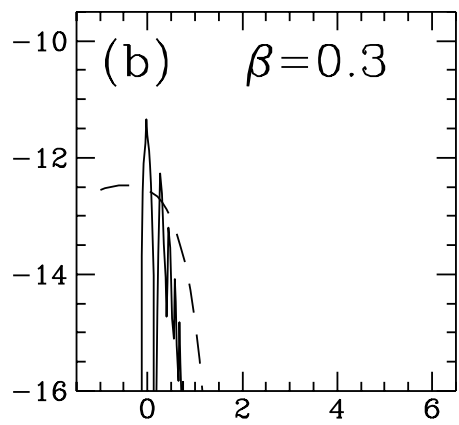
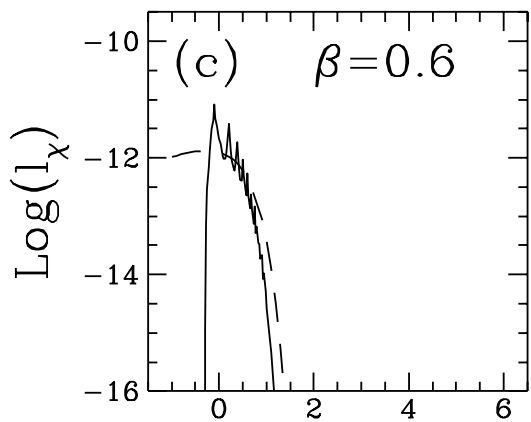
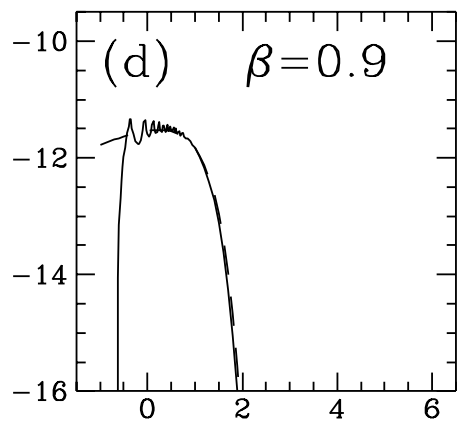
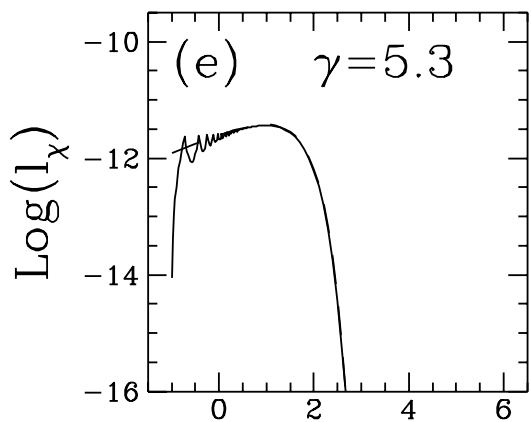
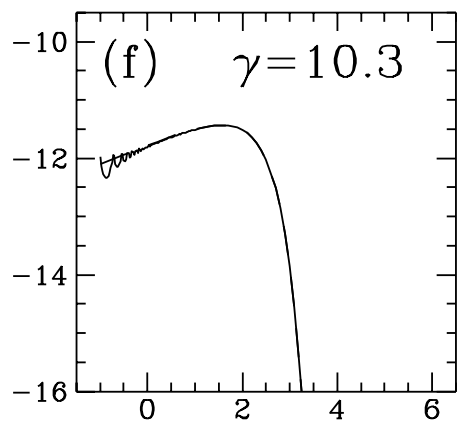
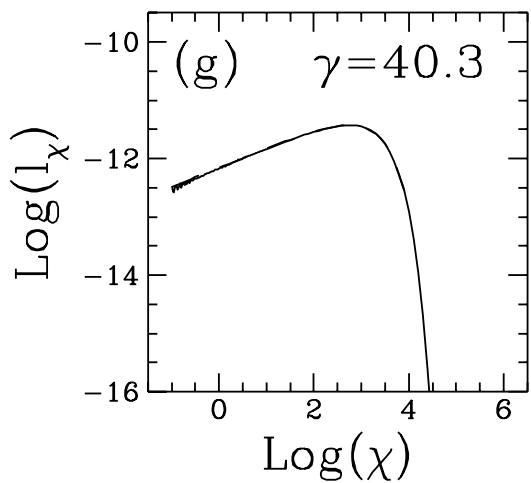
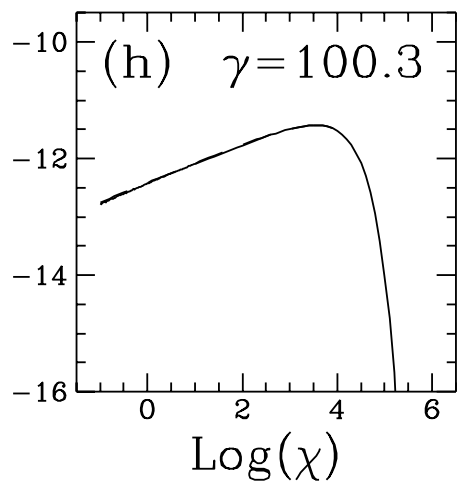
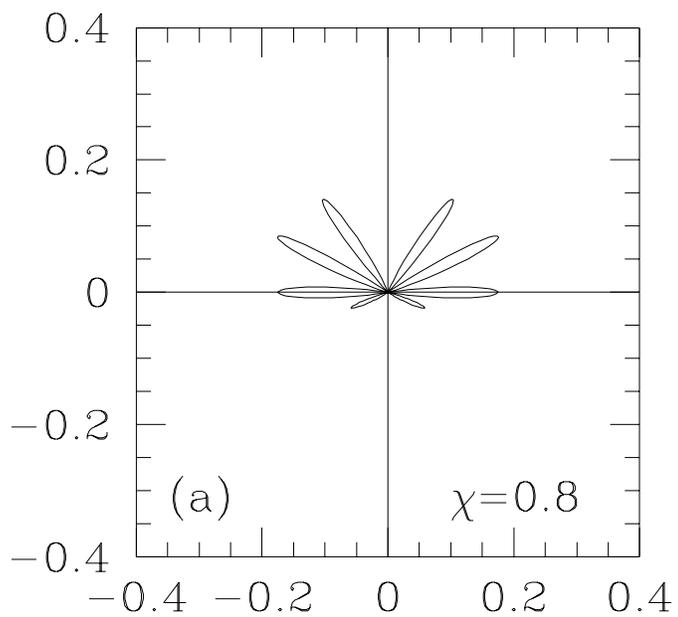
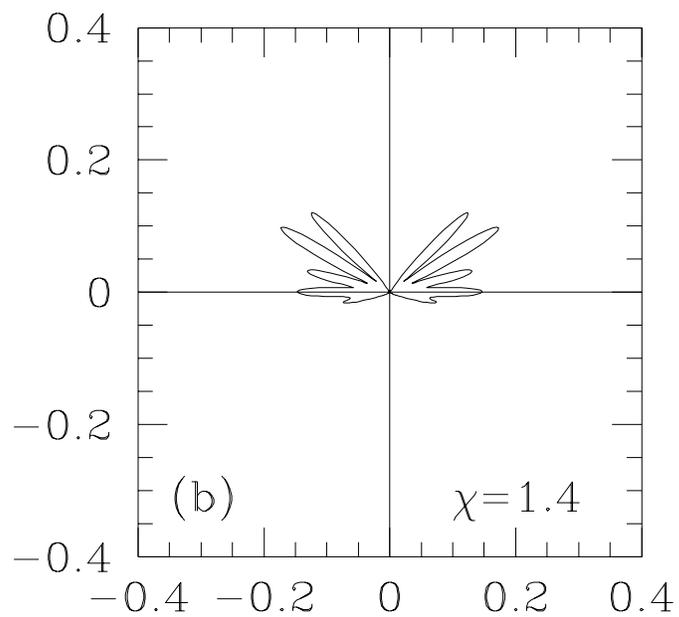
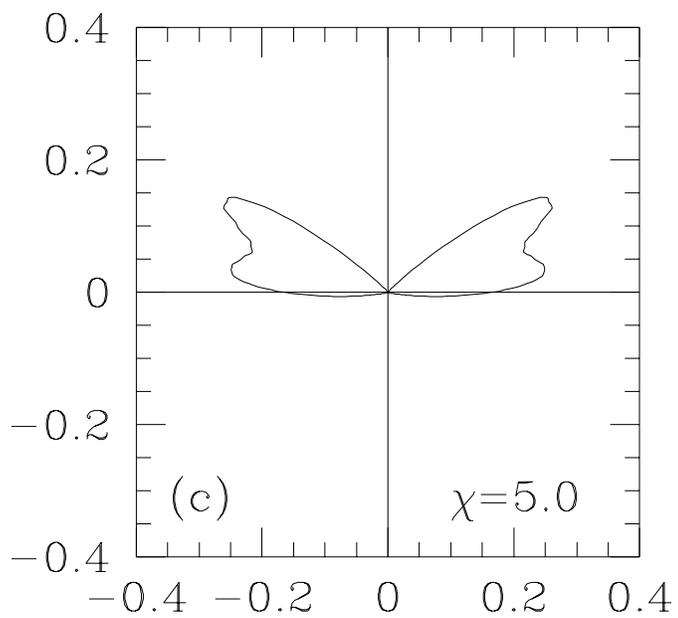
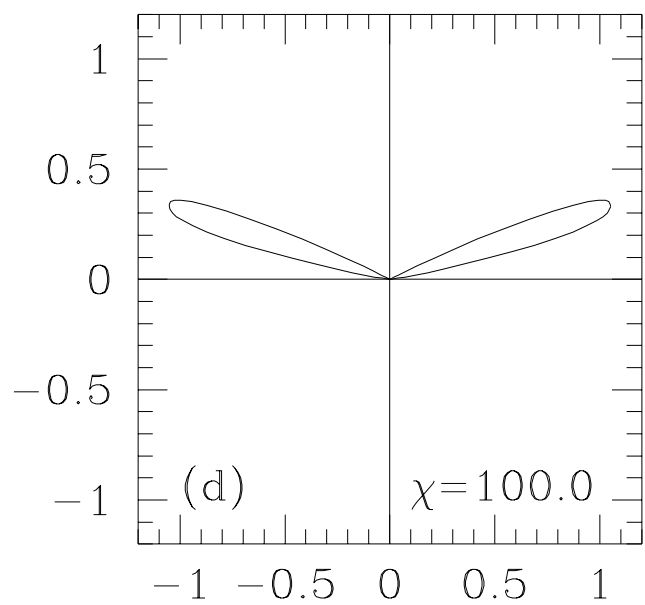
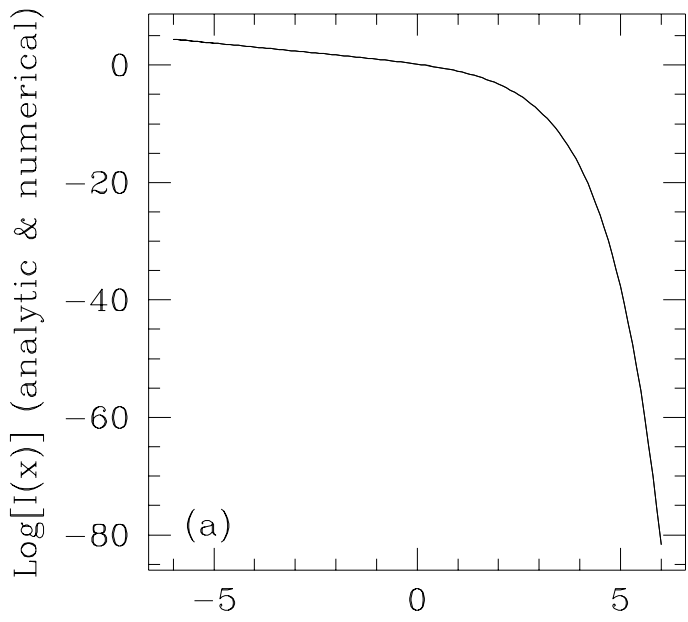
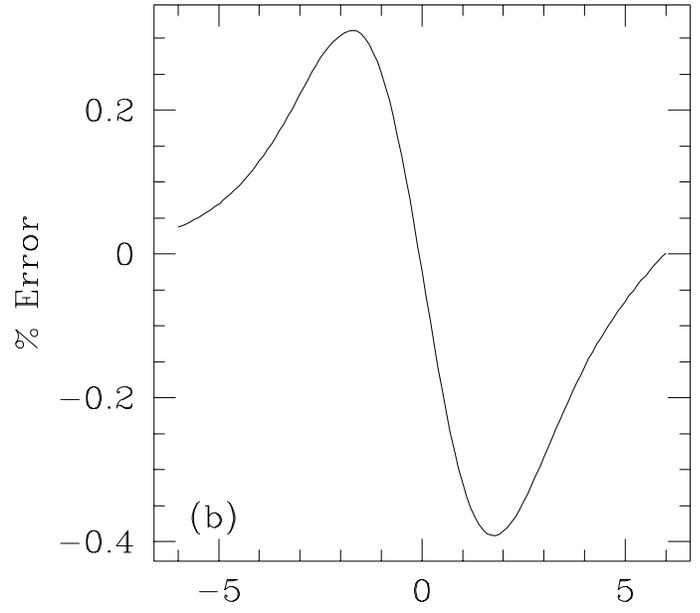
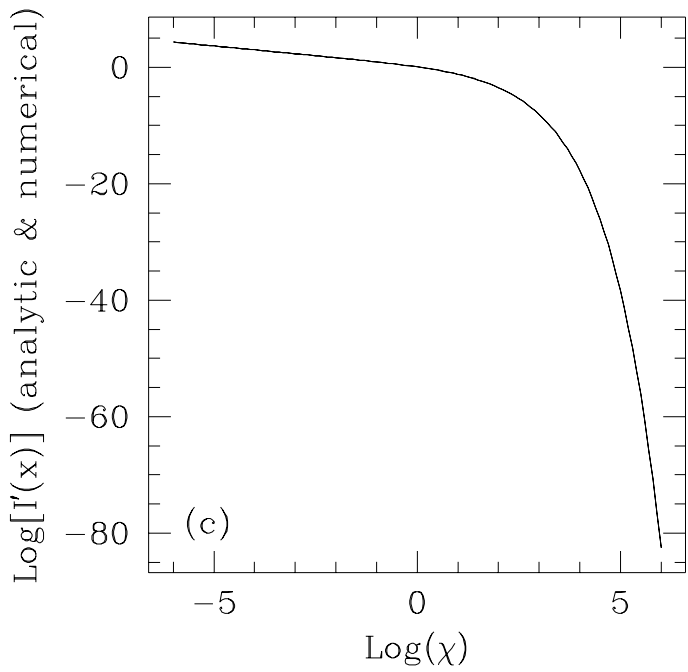
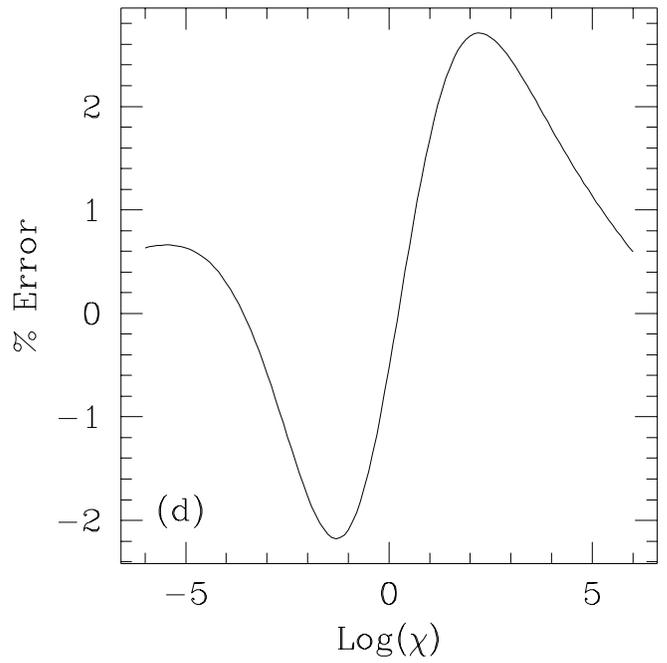
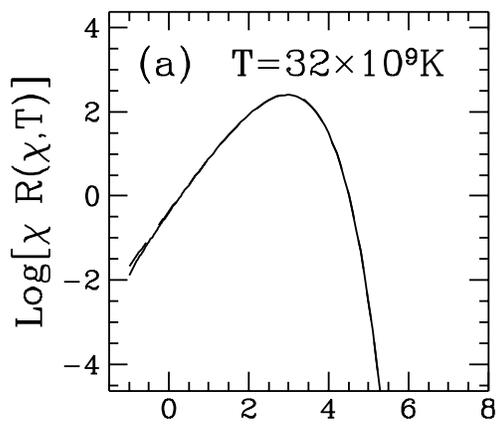
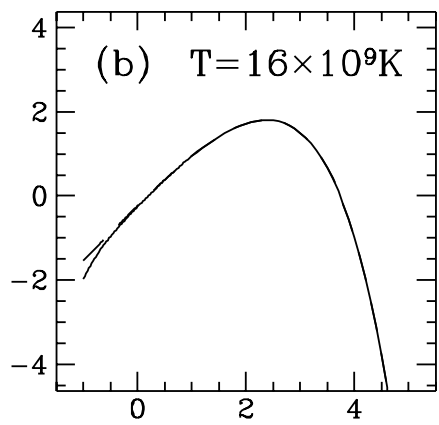
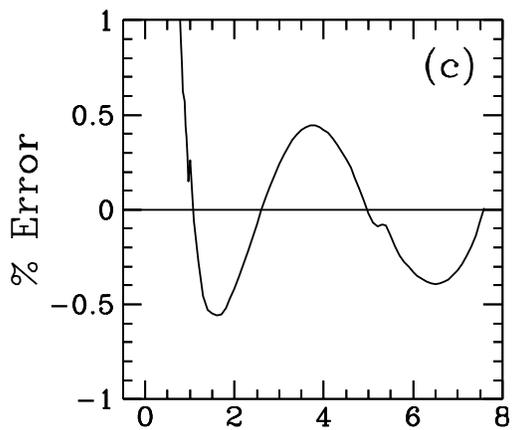
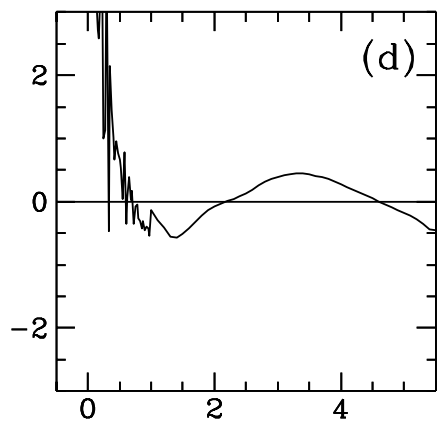
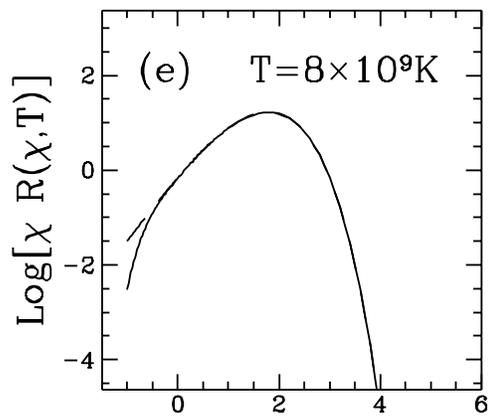
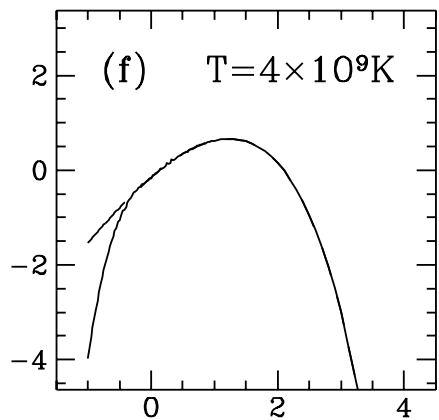
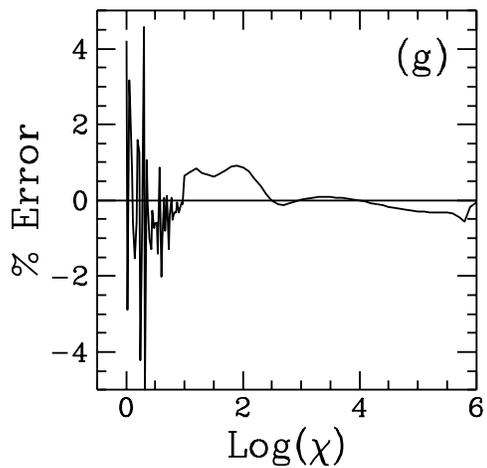
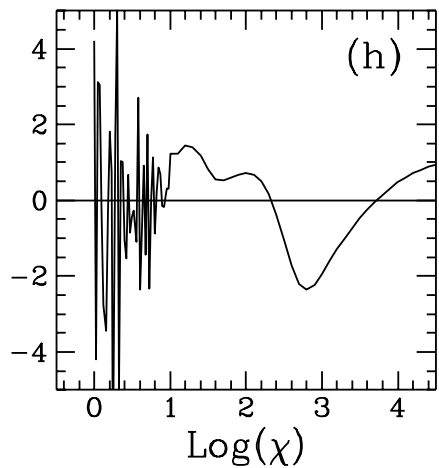
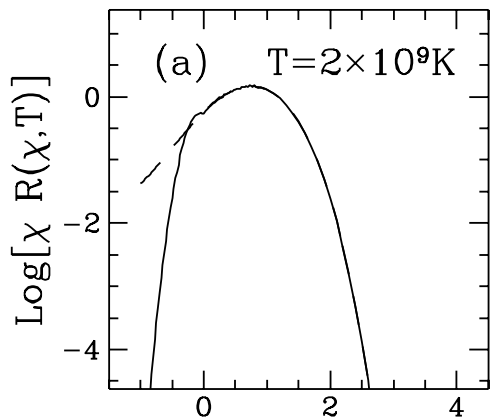
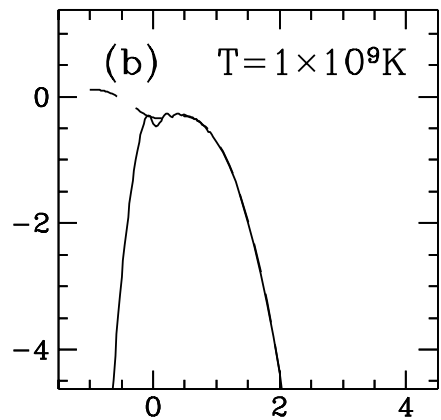
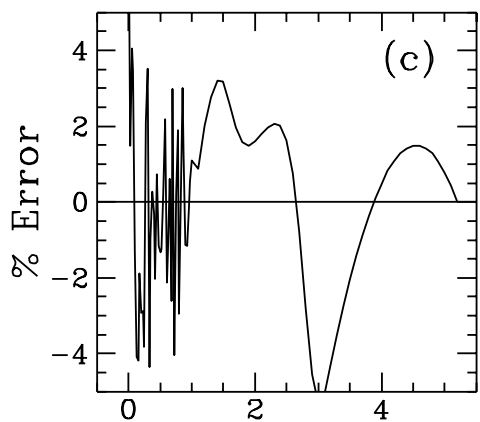
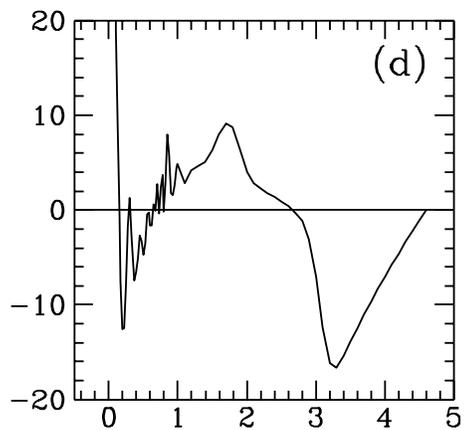
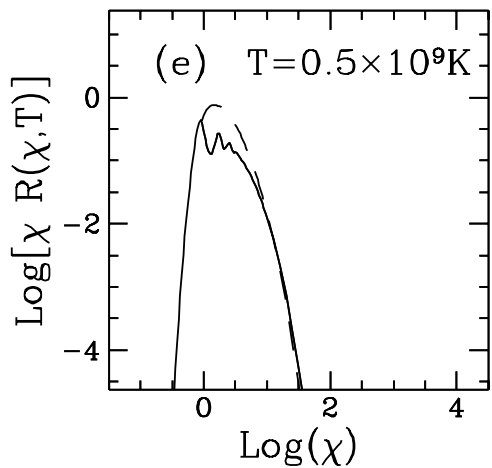
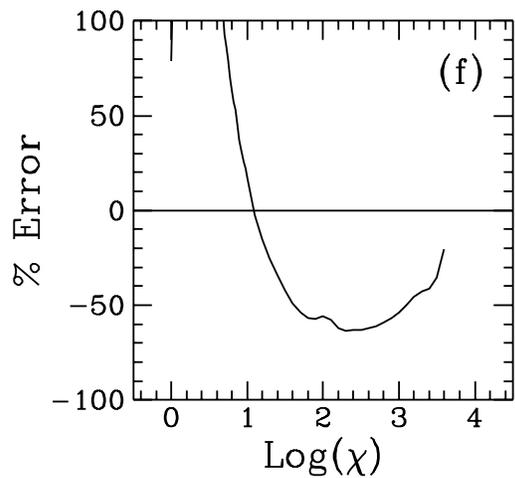
\end{document}